\documentclass{appolb}
\usepackage{epsfig}

\begin{document}
\title{ Nuclear matter properties with the re-evaluated coefficients of liquid drop model }
\author{ P. Roy Chowdhury $^*$ \thanks{E-mail:partha.roychowdhury@saha.ac.in}
\address{Saha Institute of Nuclear Physics, 1/AF Bidhan Nagar, Kolkata 700 064, India}
\and 
D.N. Basu
\address{Variable Energy Cyclotron Centre, 1/AF Bidhan Nagar, Kolkata 700 064, India}}
\maketitle
\begin{abstract}

      The coefficients of the volume, surface, coulomb, asymmetry and pairing energy terms of the semiempirical  liquid drop model mass formula have been determined by furnishing best fit to the observed mass excesses. Slightly different sets of the weighting parameters for liquid drop model mass formula have been obtained from minimizations of chisquare and mean square deviation. The most recent experimental and estimated mass excesses from Audi-Wapstra-Thibault atomic mass table have been used for the least square fitting procedure. Equation of state, nuclear incompressibility, nuclear mean free path and the most stable nuclei for corresponding atomic numbers, all are in good agreement with the experimental results.

 
\end{abstract}





\section{Introduction}
\label{section1}

      A mass formula is not only useful to the evaluation of atomic masses and nuclear binding energies but provides theoretical predictions concerning a number of features of nuclei and their behaviour. Mass models started out as empirical formulas and have now evolved into more or less full scale nuclear models, many capable of predicting a whole host of nuclear properties besides the mass. The atomic or nuclear mass is one of the most decisive factors governing nuclear stability. Enjoying reasonable success and relative ease of use are the class of mesoscopic models, where a gross macroscopic component is simply dependent on the number of nucleons and is finely sculpted using microscopic corrections for shell behaviour, deformation, nuclear incompressibility and nucleon pairing. This approach finds its orgin in the famous Bethe \cite{r1}-Weizs\"acker \cite{r2} mass formula which is an empirically refined form of the liquid drop model for the binding energy of nuclei. Since Bethe-Weizs\"acker mass formula is based on a liquid drop description of the nucleus, it is expected to reproduce the gross features of nuclear binding energies better for medium and heavy nuclei than for light nuclei. 

      Different approaches to the evaluation of the coefficients of the volume, surface, coulomb, asymmetry and pairing energy terms of the semiempirical Bethe-Weizs\"acker mass formula furnishing the best fit to the observed masses, have yielded different sets of results. Some of which are compared in Refs. \cite{r3,r4}. The mass of an atom and especially of a nucleus is not merely the sum of the masses of its constituents but is accompanied by a binding energy. In nuclear physics accurate mass measurements have a great influence as the binding energy reveals a host of nuclear structure effects, magic configuration number effects known as shell and subshell closures, shape and deformation effects and also nucleon pairing. Finally, as the binding energy determines how much energy is available for a given nuclear reaction, the impact of masses in nuclear astrophysics is also far reaching. A host of quantities derived from the masses are in play, notably, neutron and proton separation energies. This atomic mass evaluation currently containing almost harmoniously adjusted values is published every few years and the recent one is available in reference \cite{r5}. In the present work, a five parameter least square fit to the Bethe-Weizs\"acker mass formula has been achieved for new evaluations of energy coefficients by minimizing the  chisquare and the mean square deviation for atomic mass excesses using the latest compilation \cite{r5} and its consequences on nuclear properties such as incompressibility of nuclear matter, Coulomb radius constant and effect of asymmetry energy coefficient have been investigated.

\section{The liquid drop model nuclear binding energy and the atomic mass excess}
\label{section2}

      The Bethe-Weizsa\"cker formula \cite{r1,r2} is an empirically refined form of the liquid drop model for the binding energy of nuclei. Expressed in terms of the mass number A and the atomic number Z for a nucleus, the binding energy B(A,Z) is given by,

\begin{equation}
 B(A,Z) = a_vA-a_sA^{2/3}-a_cZ(Z-1)/A^{1/3}-a_{sym}(A-2Z)^2/A+\delta,
\label{seqn1}
\end{equation}
\noindent 

\begin{eqnarray}
{\rm where}~\delta=&&a_pA^{-1/2}~for~even~N-even~Z,\nonumber\\
         =&&-a_pA^{-1/2}~for~ odd~N-odd~Z,\nonumber\\
         =&&0~for~odd~A,
\label{seqn2}
\end{eqnarray} 
\noindent
and N is the neutron number of the nucleus.

      The above formula for nuclear binding energy was prescribed by Bethe \cite{r1} and Weizs\"acker \cite{r2} where the nucleus was considered as a droplet of incompressible matter and all nuclei have the same density. Scattering experiments suggest that nuclei have approximately constant density, so that the nuclear radius can be calculated by using that density as if the nucleus were a drop of a uniform liquid. Thus the first approximation to the binding energy was identified as due to the saturated exchange force and the term $a_vA$ was called the volume energy. The second term $-a_sA^{2/3}$, called the surface energy, was a correction to it since a deficit of binding energy for surface nucleons is expected due to those nucleons which are visualized as being at the nuclear surface have fewer near neighbours than which are deep within the nuclear matter. Only well known infinite range force in a nucleus is the Coulomb repulsion among protons that gives rise to the repulsive Coulomb energy term $-a_cZ(Z-1)/A^{1/3}$. This term was derived by calculating the Coulomb energy assuming that the total nuclear charge $Ze$ to be spread uniformly throughout the spherical nuclear volume of radius equal to $r_0 A^{1/3}$, $r_0$ being the Coulomb radius constant. The term $Z(Z-1)$ instead of $Z^2$ appears due to the fact that the nuclear charges are integral multiples of electronic charge $e$ and the single proton should not have any contribution to coulomb energy and hence Coulomb self energy contribution by each of Z protons should be removed. The Coulomb energy coefficient $a_c$ is given by 

\begin{equation}
 a_c = (3e^2/5r_0).
\label{seqn3}
\end{equation}
\noindent
Another deficit of binding energy depends on the neutron excess $N-Z=A-2Z$ since these excess neutrons have to go into previously unoccupied quantum states having larger kinetic energy and smaller potential energy than those already occupied. This asymmetry energy term $-a_{sym}\frac{(A-2Z)^2}{A}$, thus, arises due to purely quantum mechanical effect. The facts that there is a finite pairing energy between odd-$A$ and even-$A$ nuclei and the anomalously large binding energy for nuclei containing a magic number of neutrons and protons, fail to appear in liquid drop model since intrinsic spins of the nucleons have been omitted. To correct for this omission the pairing energy term $\delta$ was empirically incorporated.     

      Using the definition of nuclear binding energy B(A,Z) which is defined as the energy required to separate all the nucleons consituting a nucleus, the mass equation can be expressed  as 

\begin{equation}
 M_{nucleus}(A,Z) = Z m_p + (A-Z) m_n - B(A,Z),
\label{seqn4}
\end{equation}
\noindent
where $m_p$ and $m_n$ are the masses of proton and neutron respectively and $M_{nucleus}(A,Z)$ is the actual mass of the nucleus.

      The reason that the atomic mass is considered rather than the nuclear mass is that historically, the former has been the actual experimentally measured quantity, whereas  the latter is less accurate bacause its extraction requires a knowledge of binding energy of the Z atomic electrons. However, recent developments now allow the nuclear masses to be measured directly \cite{r6}. For those applications where it is necessary to know the  actual mass of the nucleus $M_{nucleus}(A,Z)$ itself, its value (in $MeV$) can be found from Eq.(4) using the nuclear binding energy or from the atomic mass excess by use of the relationship

\begin{equation}
 M_{nucleus}(A,Z) = A u + \Delta M_{A,Z} - Zm_e  + a_{el} Z^{2.39} + b_{el} Z^{5.35},
\label{seqn5}
\end{equation}   
\noindent
where $m_e$ is the mass of an electron, the atomic mass unit u is 1/12 the mass of $^{12}C$ atom, $\Delta M_{A,Z}$ is the atomic mass excess of an atom of mass number A and atomic number Z and the electronic binding energy constants \cite{r7} $a_{el} = 1.44381 \times 10^{-5}$ MeV and $b_{el} = 1.55468 \times 10^{-12}$ MeV. Hence from Eq.(4) and Eq.(5) the atomic mass excess is given by 

\begin{equation}
 \Delta M_{A,Z} = Z \Delta m_H  + (A-Z) \Delta m_n - a_{el} Z^{2.39} - b_{el} Z^{5.35} - B(A,Z)
\label{seqn6}
\end{equation}
\noindent
where $\Delta m_H = m_p + m_e - u$ =  7.28897050 + $a_{el}$ + $b_{el}$ MeV and $\Delta m_n = m_n - u$ = 8.07131710 MeV.

\section{The mass fitting and the extraction of energy coefficients}
\label{section3}

      Different approaches to the evaluation of the weighting parameters $a_v$, $a_s$, $a_c$, $a_{sym}$ and $a_p$ furnishing the best fit to the observed masses when inserted in the mass formula have yielded different sets of results. Some of which are compared in Refs. \cite{r3,r4}. The two quantities of importance that can be used for least square fitting are the root mean square deviation $\sigma$ and $\chi^2/N$ which are defined as 

\begin{equation}
 \sigma^2 =  (1/N) \sum [ \Delta M_{Th} - \Delta M_{Ex} ]^2
\label{seqn7}
\end{equation}   
\noindent

\begin{equation}
{\rm and~~} \chi^2/N =  (1/N) \sum [ ( \Delta M_{Th} - \Delta M_{Ex} ) / \Delta M_{Ex} ]^2
\label{seqn8}
\end{equation}  
\noindent
respectively where the summations extend to $N$ data points for which measured atomic mass excesses and corresponding binding energies are known. 

\begin{table}[htbp]
\caption{Energy coefficients for the Bethe-Weizs\"acker mass formula.}
\centering
\begin{tabular}{|c|c|c|c|c|}
\hline
$a_v$ & $a_s$ & $a_c$ & $a_{sym}$ & $a_p$   \\
$ MeV$ & $MeV$ & $ MeV$ & $MeV$ & $ MeV$  \\ \hline
15.260&16.267&0.689&22.209&10.076 \\
$\pm$0.020&$\pm$0.062&$\pm$0.001&$\pm$0.048&$\pm$0.854\\ \hline
$[a]$&Minimized&quantity&$\sigma^2$=&15.15 \\ \hline
&&&& \\
15.777&18.340&0.710&23.210&11.998\\
$\pm$0.053&$\pm$0.174&$\pm$0.003&$\pm$0.103&$\pm$1.912\\ \hline
$[a]$&Minimized&quantity&$\chi^2/N$=&3.336 \\ \hline
&&&& \\
15.409&16.873& 0.695&22.435&11.155\\
$\pm$0.026&$\pm$0.080&$\pm$0.002&$\pm$0.065&$\pm$0.864\\ \hline
$[b]$&Minimized&quantity&$\sigma^2$=&11.74 \\ \hline
&&&& \\
15.777&18.341&0.710&23.211&11.996\\ 
$\pm$0.037&$\pm$0.133&$\pm$0.002&$\pm$0.060&$\pm$1.536\\ \hline
$[b]$&Minimized&quantity&$\chi^2/N$=&2.484 \\ \hline
\end{tabular} 
\vskip 0.2cm
$[a]$. Using both measured 2228+extrapolated 951 atomic mass excesses.\\         
$[b]$. Using only the experimentally measured 2228 atomic mass excesses. \\

\end{table}
\noindent
The quantity $\Delta M_{Th}$ is theoretically calculated atomic mass excess obtained using Eq.(6) while $\Delta M_{Ex}$ is the corresponding experimental atomic mass excess obtained from the Audi-Wapstra-Thibault atomic mass table \cite{r5}. For the calculations of $\chi^2/N$ and the mean square deviation $\sigma^2$, masses \cite{r5} of 3179 nuclei including the 951 extrapolated values which are predicted according to systematics, have been used for least square fitting. These calculations have been repeated using only the 2228 experimental data. The $\chi^2/N$ and $\sigma^2$ minimizations have been performed separately since both can not be minimized with identical sets of values for energy coefficients. In case of the $\chi^2/N$ minimization, the denominator has to be kept equal to unity only for $^{12}C$ in order to accommodate it in the calculation as the atomic mass excess of $^{12}C$ is, by definition, equal to zero. In table-1, values of the energy coefficients along with the minimized quantities with their values have been listed. Since Bethe-Weizs\"acker mass formula is more appropriate for nuclear binding energies of medium and heavy nuclei than for light nuclei, the values obtained for $\sigma^2$ and $\chi^2/N$ are not very small. However, as can be seen from table-1, the errors associated with the energy coefficients $a_v$, $a_s$, $a_c$ and $a_{sym}$ are very small while that with $a_p$ is reasonably small. 

      As can be seen from the table-1, both $\sigma^2$, which is a measure of absolute error, and the $\chi^2/N$, which is a measure of relative error, can not be minimized simultaneously with the same set of parameters. The set of parameters obtained by minimizing $\chi^2/N$ are almost same as those listed in Refs. \cite{r8,r9}. However, in many cases, such as reaction Q value calculations, absolute errors involved in atomic mass excesses are the  quantities of concern and $\sigma^2$ minimization plays more important role than $\chi^2/N$ minimization. On the contrary, $\chi^2/N$ minimization causes minimization of relative, that is, uniform percentage error involved in the mass predictions.

\section{Energy coefficients and the nuclear matter properties }
\label{section4}

      A density dependent M3Y effective nucleon-nucleon (NN) interaction based on the G-matrix elements of the Reid-Elliott NN potential has been used to determine the nuclear matter equation of state. The equilibrium density of the nuclear matter has been determined by minimizing the energy per nucleon. The density dependence parameters have been chosen to reproduce the saturation energy per nucleon and the saturation density of spin and isospin symmetric cold infinite nuclear matter. The general expression for the density dependent effective NN interaction potential $v(s)$ is written as  

\begin{equation}
 v(s,\rho, \epsilon) = t^{M3Y}(s, \epsilon) g(\rho, \epsilon)~MeV
\label{seqn9}
\end{equation}   
\noindent
where M3Y effective interaction potential supplemented by a zero range pseudopotential $t^{M3Y}$ is given by  

\begin{equation}
 t^{M3Y}(s,\epsilon) = 7999~(MeV).~\frac{e^{- 4s}}{ 4s} - 2134~(MeV).~\frac{e^{- 2.5s}}{2.5s} + J_{00}(\epsilon) \delta(s)
\label{seqn10}
\end{equation}   
\noindent
where the costants 4 and 2.5 have dimensions of $fm^{-1}$, $\delta(s)$ has the dimension of $fm^{-3}$ and the zero-range pseudo-potential $J_{00}(\epsilon)$ representing the single-nucleon exchange term is given by 

\begin{equation}
 J_{00}(\epsilon) = -276 (1 - \alpha\epsilon) (MeV.fm^3)
\label{seqn11}
\end{equation}   
\noindent
and the dimensionless density dependent part is given by  

\begin{equation}
 g(\rho, \epsilon) = C (1 - \beta(\epsilon)\rho^{2/3}) 
\label{seqn12}
\end{equation}   
\noindent

      The energy per nucleon $\epsilon$ obtained using the effective nucleon-nucleon interaction $v(s)$ for the spin and isospin symmetric cold infinite nuclear matter, which will henceforth be called the standard nuclear matter, is given by

\begin{equation}
 \epsilon = [3\hbar^2k_F^2/10m] + g(\rho, \epsilon)\rho J_v / 2 = [3\hbar^2k_F^2/10m] + [\rho J_v C (1 - \beta\rho^{2/3})/2]
\label{seqn13}
\end{equation}   
\noindent
where m is the nucleonic mass, $k_F=(1.5\pi^2\rho)^{1/3}$ is the Fermi momentum, $\rho$ is the nucleonic density while $\rho_{0}$ being the saturation density for the standard nuclear matter and $J_v$ represents the volume integral of $t^{M3Y}$, the M3Y interaction supplemented by the zero-range pseudopotential. The Eq.(13) can be  differentiated with respect to $\rho$ to yield equation  

\begin{equation}
 \partial\epsilon/\partial\rho = [\hbar^2k_F^2/5m\rho] + J_v C [1 - (5/3)\beta\rho^{2/3}] /2 
\label{seqn14}
\end{equation}
\noindent
The equilibrium density of the nuclear matter is determined from the saturation condition $\partial\epsilon/\partial\rho = 0$. Then Eq.(13) and Eq.(14) with the saturation condition can be solved simultaneously for fixed values of the saturation energy per nucleon $\epsilon_0$ and the saturation density $\rho_{0}$ of the standard nuclear matter, to obtain the values of the density dependence parameters $\beta$ and C given by

\begin{equation}
 \beta = [(1-p)\rho_{0}^{-2/3}]/[3-(5/3)p],
\label{seqn15}
\end{equation} 
\noindent

\begin{equation}
 p = [10m\epsilon_0]/[\hbar^2k_{F_0}^2],
\label{seqn16}
\end{equation} 
\noindent
 
\begin{equation}
 k_{F_0} = [1.5\pi^2\rho_0]^{1/3},
\label{seqn17}
\end{equation} 
\noindent

\begin{equation}
 C = -[2\hbar^2k_{F_0}^2] / [5mJ_v\rho_0(1 - (5/3)\beta\rho_0^{2/3})],
\label{seqn18}
\end{equation} 
\noindent
respectively. It is quite obvious that the density dependence parameter $\beta$ obtained by this method depends only on the saturation energy per nucleon $\epsilon_0$, the saturation density $\rho_{0}$ but not on the parameters of the M3Y interaction while the other density dependence parameter C depends on the parameters of the M3Y interaction also through the volume integral $J_v$. The energy per nucleon can be rewritten as    
    
\begin{equation}
 \epsilon = [3\hbar^2k_F^2/10m] - (\rho/\rho_{0}) [\hbar^2k_{F_0}^2 (1 - \beta\rho^{2/3})]/[5m(1 - (5/3)\beta\rho_{0}^{2/3})]
\label{seqn19}
\end{equation}
\noindent
and the incompressibility $K_0$ of the spin and isospin symmetric cold infinite nuclear matter is given by  
  
\begin{equation}
 K_0 = k_F^2\partial^2\epsilon/\partial{k_F^2} = 9\rho^2\partial^2\epsilon/\partial\rho^2\mid_{\rho=\rho_0} = [-(3\hbar^2k_{F_0}^2/5m) - 5 J_v C \beta\rho_0^{5/3}]
\label{seqn20}
\end{equation}
\noindent
The pressure $P$ and the energy density $\varepsilon$ of nuclear matter can be given by 

\begin{equation}
 P = \rho^2 \partial\epsilon/\partial\rho = [\rho \hbar^2k_F^2/5m] + \rho^2 J_v C [1 - (5/3) \beta\rho^{2/3}]/2,
\label{seqn21}
\end{equation} 
\noindent

\begin{equation}
 \varepsilon = \rho (\epsilon + m c^2) = \rho [(3\hbar^2k_F^2/10m) + \rho J_v C (1 - \beta\rho^{2/3})/2 + m c^2], 
\label{seqn22}
\end{equation} 
\noindent
respectively, and thus the velocity of sound $v_s$ in standard nuclear matter is given by 

\begin{equation}
 \frac{v_s}{c}=\sqrt{ \frac{\partial P}{\partial\varepsilon}} =\sqrt{[2\rho\frac{\partial\epsilon}{\partial\rho}-\frac{\hbar^2k_F^2}{15m}- \frac{5}{9}J_vC\beta \rho^{5/3}]/[\epsilon + m c^2 + \rho \frac{\partial\epsilon}{\partial\rho}]} \nonumber\\
\label{seqn23}
\end{equation} 
\noindent

\begin{figure}[htbp]
\eject\centerline{\epsfig{file=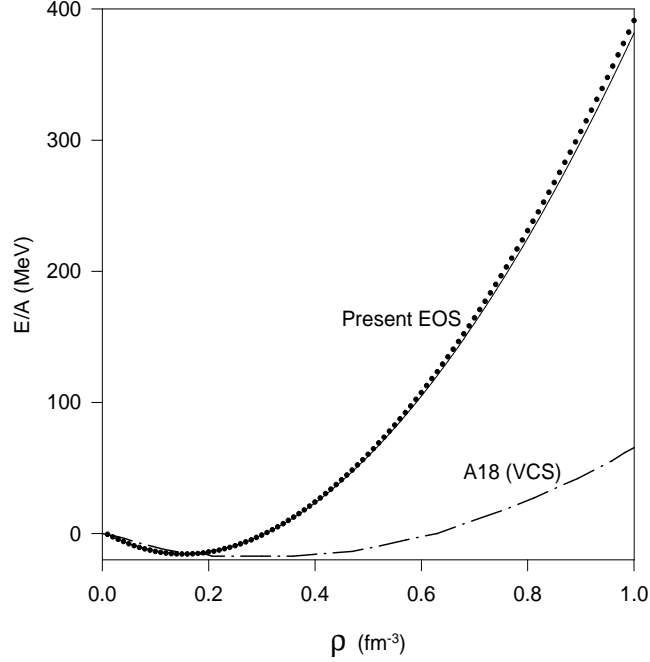,height=8.9cm,width=8.9cm}}
\caption
{The energy per nucleon $\epsilon$ of nuclear matter as a function of $\rho$. The continuous line represents the curve for the present calculations using saturation energy per nucleon of -15.26 MeV and the dots represent the same using saturation energy per nucleon of -15.78 MeV whereas the dash-dotted line represents the same for the A18 model using variational chain summation (VCS) [13] for the spin and isospin symmetric infinite nuclear matter.}
\label{fig1}
\end{figure}

      One of the most directly derivable information provided by the mass formula is the magnitude of the Coulomb radius constant contained within the expression for the Coulomb energy constant $ a_c = (3e^2/5r_0)$ assuming a uniform volume distribution of charge. Using the values of Coulomb energy coefficients $a_c$, the values obtained for the Coulomb radius constant $r_0$ have been listed in table-2. The present value for the Coulomb radius constant is $r_0=1.22-1.25 fm$ which like previous such mass fittings still overpredicts. The crucial property of nuclear matter is the saturation density. It has been recognised right from the beginning that Bethe-Weizs\"acker mass formula  or its improved versions can not give this property through its Coulomb radius constant using the simple relation $\rho_0=3/4\pi r_0^3$. This density is measured through electron scattering experiments on heavy nuclei. This gives a value corresponding to $r_0=1.12-1.13$ fm.  Identifying $a_v$ as the saturation energy per nucleon for the spin and the isospin symmetric cold infinite nuclear matter and hence using the saturation energy per nucleon equal to $-15.260 MeV$ along with the density dependent M3Y effective interaction and the commonly used value for saturation density equal to $0.1533 fm^{-3}$ \cite{r10}, as one can see in table-2, the nuclear incompressibility is found to be $293.4 MeV$ which is in close agreement to experimental data \cite{r11,r12}. In Fig.~\ref{fig1} the energy per nucleon E/A=$\epsilon$ of standard nuclear matter is plotted as a function of nucleonic density $\rho$. The continuous line represents the curve using saturation energy per nucleon of $-15.26 MeV$ and the dots represent the same using saturation energy per nucleon of $-15.78 MeV$ whereas the dash-dotted line represents the same for the A18 model using variational chain summation (VCS) \cite{r13} for the spin and isospin symmetric infinite nuclear matter. The minima of the energies per nucleon equaling the saturation energies of -15.260 MeV and -15.777 MeV for the present calculations occur correctly at the saturation density $\rho_0=0.1533 fm^{-3}$ while that for the A18(VCS) model occurs around $\rho=0.28 fm^{-3}$ with a saturation energy of about -17.3 MeV. 

\begin{table}[htbp]
\caption{Coulomb radius constant and the compression modulus using the energy dependence parameter $\alpha=0.005 MeV^{-1}$ and values of the saturation energy per nucleon $\epsilon_0$ determined by $a_v$ and using the saturation density $\rho_0=0.1533 fm^{-3}$ for the standard nuclear matter. Rows (a) and (b) correspond to $a_v$ and $a_c$ values obtained from minimizing $\sigma^2$ and $\chi^2/N$ respectively using both measured 2228+extrapolated 951 atomic mass excesses whereas rows (c) and (d) correspond to $a_v$ and $a_c$ values obtained from minimizing $\sigma^2$ and $\chi^2/N$ respectively using only the experimentally measured 2228 atomic mass excesses.} 
\centering
\begin{tabular}{|c|c|c|c|c|c|c|}
\hline
&$\epsilon_0$ & $a_c$ & $r_0$ & $\beta$  & C & $K_0$      \\
&$MeV$ & $MeV$ &$fm$&$fm^2$&  & $MeV$   \\ \hline
 (a)&15.260&0.689&1.254&1.668&2.07 &293.4\\ \hline
 (b)&15.777&0.710&1.217&1.676&2.11&301.1 \\ \hline
 (c)&15.409&0.695&1.244&1.671&2.08 &295.6\\ \hline
 (d)&15.777&0.710&1.217&1.676&2.11&301.1 \\ \hline 
\end{tabular} 
\end{table}

      In table-3 the theoretical estimates of the pressure $P$ and velocity of sound $v_s$ of standard nuclear matter have been listed as functions of nucleonic density $\rho$ and energy density $\varepsilon$ using the usual value of 0.005/MeV for the parameter $\alpha$ of energy dependence, given in eqn.(11), of the zero range pseudo-potential. As for any other non-relativistic EOS, present EOS also suffers from superluminosity at very high densities. According to present calculations the velocity of sound becomes imaginary for $\rho\le 0.1fm^{-3}$ and exceeds the velocity of light c at $\rho \ge5.3\rho_0$ and the EOS obtained using $v_{14}+TNI$ \cite{r14} also resulted in sound velocity becoming imaginary at same nuclear density and  superluminous at about the same nuclear density. But in contrast, the incompressibility $K_0$ of infinite nuclear matter for the $v_{14}+TNI$ was chosen to be 240 MeV while that obtained by the present theoretical estimate is 290-300 MeV which is in excellent agreement with the experimental value of $K_0=300\pm25$ MeV obtained from the giant monopole resonance \cite{r11} and with the the recent experimental determination of $K_0$ based upon the production of hard photons in heavy ion collisions which led to the experimental estimate of $K_0=290\pm50$ MeV \cite{r12}. 

\begin{table}[htbp]
\caption{Energy per nucleon $\epsilon$, pressure $P$, energy density $\varepsilon$ and velocity of sound $v_s$ as  functions of nuclear density $\rho$ using saturation energy per nucleon equal to $-15.260 MeV$ and saturation density = $0.1533 fm^{-3}$ for standard nuclear matter.}
\centering
\begin{tabular}{|c|c|c|c|c|c|}
\hline
$\rho$&$\rho/\rho_{0}$&$\epsilon$&P&$\varepsilon $&$v_s$  \\
$fm^{-3}$& &$MeV$   &$MeV fm^{-3}$&$MeV fm^{-3}$&in units of c       \\ \hline
     .01&  .6523E-01& -.7537E+00& -.1677E-01&  .9382E+01&  .0000E+00\\ 
     .10&  .6523E+00& -.1325E+02& -.7633E+00&  .9257E+02&  .0000E+00\\
     .20&  .1305E+01& -.1378E+02&  .2520E+01&  .1850E+03&  .2879E+00\\
     .30&  .1957E+01& -.1138E+01&  .1689E+02&  .2813E+03&  .4700E+00\\
     .40&  .2609E+01&  .2341E+02&  .4829E+02&  .3849E+03&  .6207E+00\\ 
     .50&  .3262E+01&  .5896E+02&  .1020E+03&  .4989E+03&  .7442E+00\\
     .60&  .3914E+01&  .1048E+03&  .1830E+03&  .6262E+03&  .8443E+00\\
     .70&  .4566E+01&  .1605E+03&  .2958E+03&  .7696E+03&  .9248E+00\\
     .80&  .5219E+01&  .2254E+03&  .4447E+03&  .9315E+03&  .9895E+00\\ 
    .90&  .5871E+01&  .2993E+03&  .6340E+03&  .1114E+04&  .1042E+01\\
   1.00&  .6523E+01&  .3819E+03&  .8675E+03&  .1321E+04&  .1084E+01\\ \hline
\end{tabular} 
\end{table}

\begin{figure}[htbp]
\eject\centerline{\epsfig{file=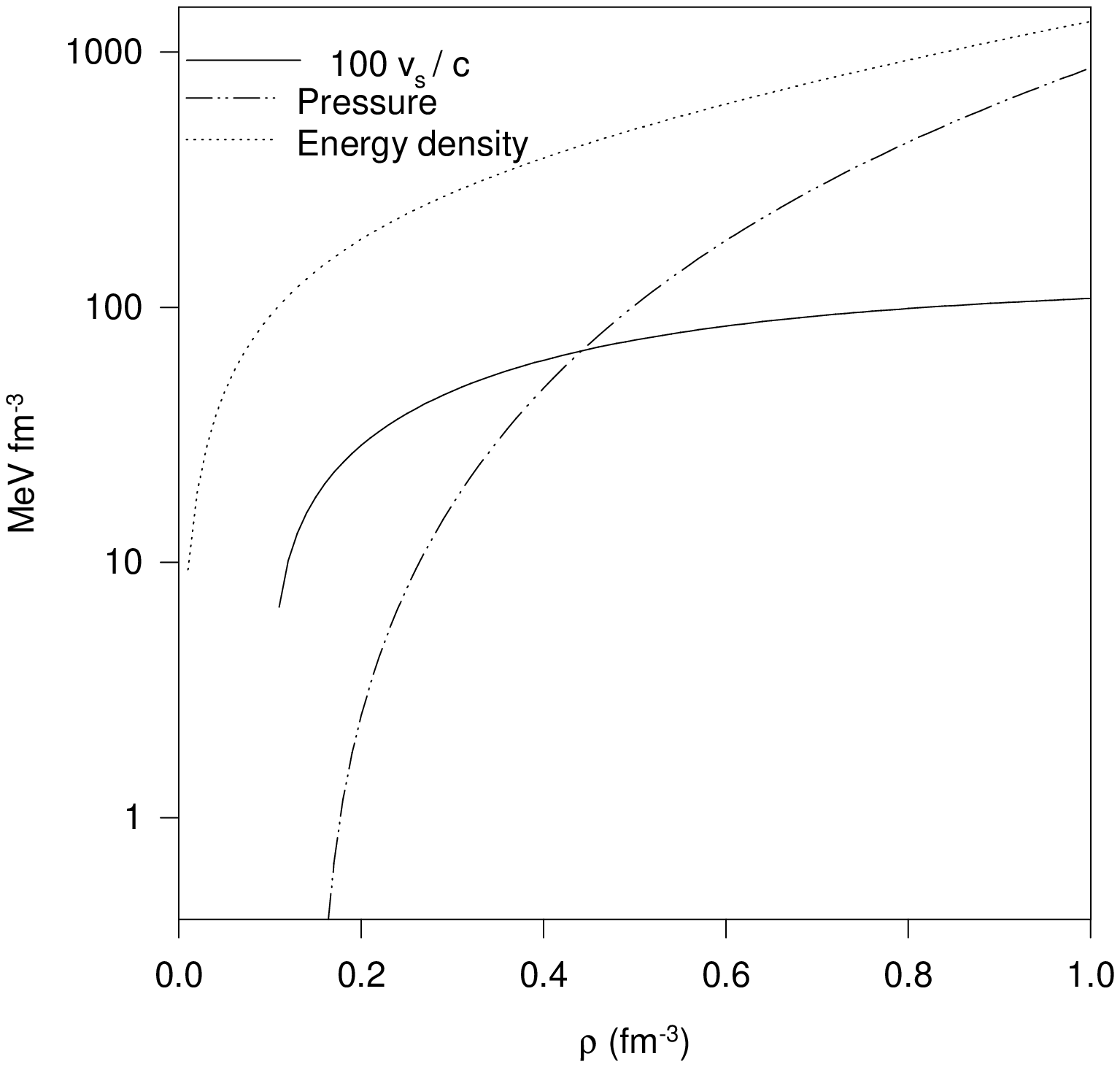,height=8.9cm,width=8.9cm}}
\caption
{The velocity of sound $v_s$ in nuclear matter, the pressure $P$ and the energy density $\varepsilon$ of nuclear matter as a function of nuleonic desity $\rho$. The continuous line represents the velocity of sound in units of $10^{-2}c$ for the standard nuclear matter, the dash-dotted line represents pressure in $MeV fm^{-3}$ while the dotted line represents energy density in $MeV fm^{-3}$ for the standard nuclear matter.}
\label{fig2}
\end{figure}

      In Fig.~\ref{fig2}, the plots of the velocity of sound $v_s$ in nuclear matter, the pressure $P$ and the energy density $\varepsilon$ of nuclear matter as functions of nuleonic density $\rho$ have been shown. The continuous line represents the velocity of sound in units of $10^{-2}c$ for the standard nuclear matter, the dash-dotted line represents  pressure in $MeV fm^{-3}$ while the dotted line represents energy density in $MeV fm^{-3}$ for the standard nuclear matter.

      Keeping A constant while differentiating Eq.(4) and using Eq.(1) and setting the term $\partial M_{nucleus}(A,Z)/\partial Z\mid_A$ equal to zero, one obtains 

\begin{equation}
  Z_{stable} = \frac{[A+(a_cA^{2/3}/2x)]}{[(4a_{sym}/x)+(a_cA^{2/3}/x)]}~{\rm where}~x=2a_{sym}+[(m_n-m_p)/2].
\label{seqn24}
\end{equation}
\noindent
The second term in the numerator of the above equation is small compared to the atomic mass number A. The above relation which connects Z with A for the most stable nuclei can be written in a closed form as 

\begin{equation}
Z_{stable} = \frac{[A+0.5a_1 A^{2/3}]}{[a_2 + a_1 A^{2/3}]}~{\rm where}~a_1=a_c/x,~{\rm and}~a_2=4a_{sym}/x.
\label{seqn25}
\end{equation}
\noindent
Table-4 provides values of $a_1$, $a_2$ for different sets of co-efficients $a_c$, $a_{sym}$. 

\begin{figure}[htbp]
\eject\centerline{\epsfig{file=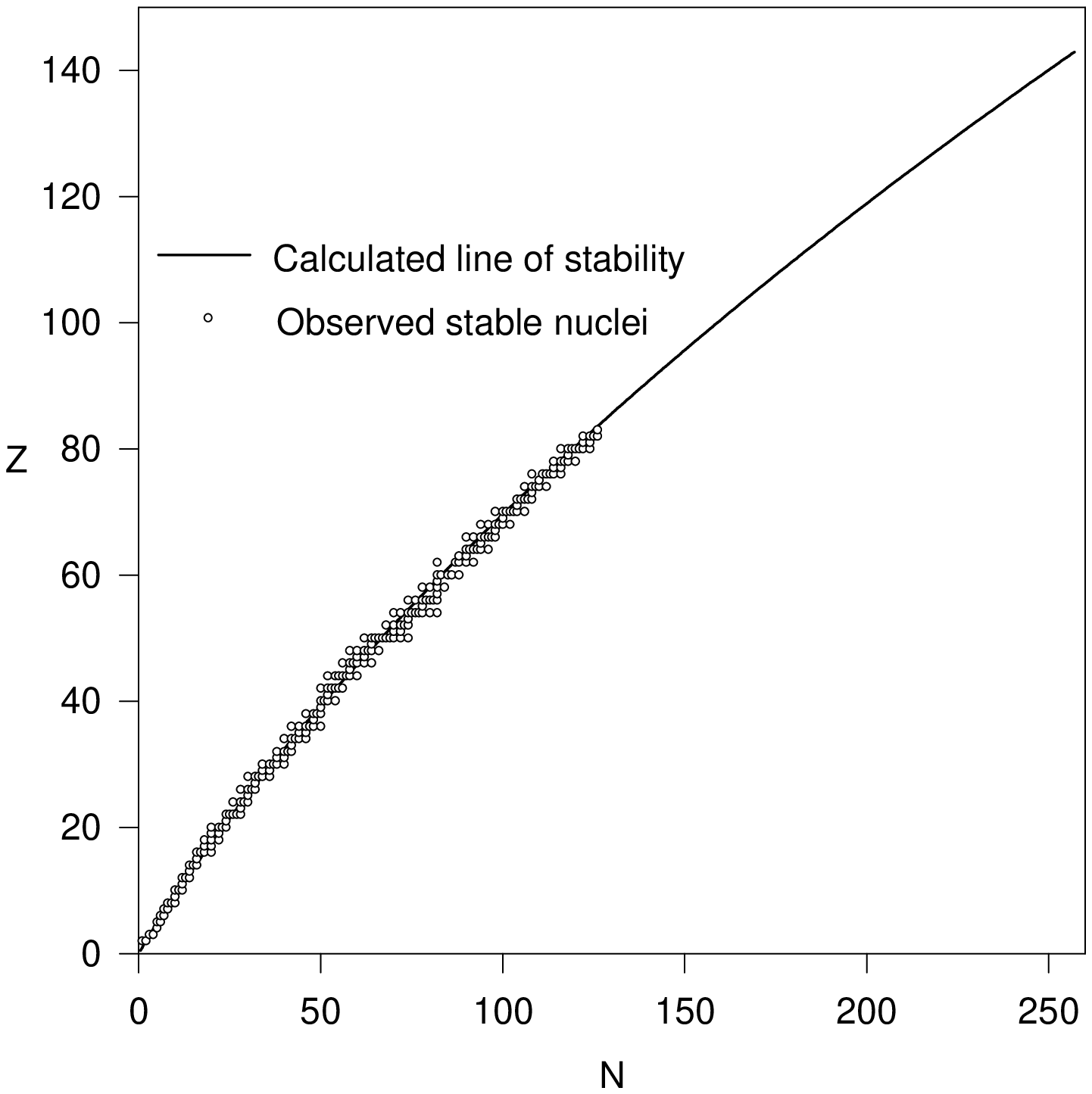,height=8.9cm,width=8.9cm}}
\caption
{The plot of Z versus N for the most stable nuclei. The continuous line represents the theoretical curve while the dots represent observed stable nuclei.}
\label{fig3}
\end{figure}

\begin{table}[htbp]
\caption{The values of $a_1$ and $a_2$. } 
\centering
\begin{tabular}{|c|c|c|c|}
\hline
$a_c$ & $a_{sym}$ & $a_1$  & $a_2$      \\
$MeV$ & $MeV$ & &  \\ \hline
 0.689&22.209&0.0153&1.971\\ \hline
 0.710&23.210&0.0151&1.973 \\ \hline
 0.695&22.435&0.0153&1.972\\ \hline
 0.710&23.211&0.0151&1.973 \\ \hline 
\end{tabular} 
\end{table}
\noindent
Since the values of $a_1$ and $a_2$ do not differ significantly for the different sets of co-efficient values $a_c$ and $a_{sym}$, in Fig.~\ref{fig3} the plot of Z versus N=A-Z for the most stable nuclei is shown for the first set of values only. The continuous line represents the theoretical curve following the exact expression given by Eq.(25) while the dots represent the observed stable nuclei from the recent nuclide chart \cite{r15}. 

      The density dependence parameter $\beta$ has the dimension of cross section. The density dependent term $(1 - \beta\rho^{2/3})$ reduces the strength of the interaction which changes sign at high densities making it repulsive. It is a direct consequence of the Pauli blocking effect. Thus $(1 - \beta\rho^{2/3})$ can be interpreted as the probability of non-interaction arising due to the collision probability $\beta\rho^{2/3}$ of a nucleon in nuclear medium of density $\rho$. The density dependence parameter $\beta$ can be identified as the in medium effective nucleon-nucleon interaction cross-section $\sigma_0$. Density dependence parameter $\beta$ along with nucleonic density of infinite nuclear matter $\rho_0$ can, therefore, provide the nuclear mean free path $\lambda=1/(\rho_0 \sigma_0)$. Using value of the density dependence parameter $\beta=1.668 fm^2$ along with the nucleonic density of $0.1533 fm^{-3}$, the value obtained for the nuclear mean free path $\lambda$ is about $3.9 fm$. 

\section{Summary and conclusion}
\label{section5}

      In summary, we conclude that five parameter least square fits to the Bethe-Weizs\"acker mass formula by minimizing $\sigma^2$ and $\chi^2/N$ yield slightly different sets of energy coefficients $a_v$, $a_s$, $a_c$, $a_{sym}$ and $a_p$. Both $\sigma^2$ and $\chi^2/N$ can not be minimized simultaneously by the identical sets of parameters. The $\sigma^2$ minimization is more appropriate since it reduces absolute errors involved in mass predictions. Identifying $a_v$ as the saturation energy per nucleon for the spin and the isospin symmetric cold infinite nuclear matter along with the value for saturation density equal to $0.1533 fm^{-3}$, the nuclear incompressibility is found to be $290-300 MeV$ which is in excellent agreement with the experimental estimates from GMR \cite{r11} as well as determination based upon the production of hard photons in heavy ion collisions \cite{r12}. The present theoretical estimate of nuclear incompressibility is in reasonably close agreement with other theoretical estimates obtained by INM \cite{r16} model, using the Seyler-Blanchard interaction \cite{r10} or the relativistic Brueckner-Hartree-Fock theory \cite{r17}. The value of $\beta$ provide nuclear mean free path $\lambda = 3.89 - 3.91 fm$ which is in excellent agreement with the values derived by other methods \cite{r18}.  



\end{document}